\documentclass[amsmath,preprint]{aastex701}

\begin{document}

\title{The nine rings of the galaxy LEDA 1313424}

\date{May 11, 2026}

\author{Pierre Sikivie}
\affiliation{Department of Physics, University of Florida,
Gainesville, FL 32611, USA}
\email{sikivie@ufl.edu}

\author{Yuxin Zhao}
\affiliation{Department of Physics, University of Florida,
Gainesville, FL 32611, USA}
\email{yuxinzhao@ufl.edu}

\begin{abstract}
The galaxy LEDA~1313424 has recently been discovered by I. Pasha et al. to have nine rings. The authors of the discovery paper interpret the rings to be the result of the passing of a smaller galaxy through the center of the LEDA galaxy 56~Myr ago. In that interpretation the outermost observed ring, which has approximately 70~kpc radius, would have traveled at 1,220~km/s, an implausibly large speed. The expected speed of the 70~kpc ring is 55~km/s in the analytic theory of collisional ring galaxies used in the discovery paper to explain the pattern of ring radii. Furthermore, the widths of the observed rings are inconsistent with the predictions of the analytic theory and the assumptions underlying the analytic theory are violated in the specific case of LEDA. As an alternative interpretation we propose that the rings of LEDA are the imprint upon baryonic matter of caustic rings of dark matter. Our proposal requires that the dark matter is axions forming a Bose-Einstein condensate.
\end{abstract}

\section{Introduction}

Beginning with the discovery of the Cartwheel galaxy by \citet{Cartwheel}, many galaxies have been observed to be surrounded by luminous rings. A catalog of ringed galaxies is provided in 
\citet{Madore}.  Most ringed galaxies have a 
single observed ring.  Some have two or three  
\citep{Struck}. Until \citet{bull}, none had 
more than three. So, regardless of the interpretation 
given, the fact that LEDA 1313424 has nine rings is 
a remarkable discovery.

The rings surrounding galaxies are commonly thought 
to be the outcome of the passing of an impact galaxy 
through the center of the ringed galaxy in a head-on 
collision.  The collision causes density waves in 
the disk of the target galaxy \citep{Lynds}.  The 
density waves may produce caustics of baryonic matter 
(stars and gas) in the disk, and the caustics are 
thought to give rise to the observed rings.  This 
interpretation is widely accepted, and ringed 
galaxies are commonly termed ``collisional ring 
galaxies'', or CRGs.  

The LEDA galaxy has redshift z = 0.039414, 
corresponding to a luminosity distance of 
173.9~Mpc \citep{bull}. It is an unusually  
large disk galaxy, about twice as large as 
the Milky Way.  Its rotation speed is 
approximately 370 km/s \citep{bull}.  Its 
nine rings appear as ellipses because 
the normal to the galactic disk, which 
contains the rings, makes an angle 
$(41.9 \pm 1.5)^\circ$ with the line of 
sight to the galaxy \citep{bull}. The 
orientations and axis ratios of the 
ellipses are consistent with the rings 
being circles in a single plane, although 
the ellipses are not exactly 
concentric \citep{bull}.
  
The first two columns of Table I give
the radii of LEDA's nine rings in units 
of pixels, and the errors on the radii.
The entries were read off from Fig.~4 of 
\citet{bull}.  The radius of the 
outermost observed ring is about 70~kpc.  Part 
of the evidence in support of the CRG
interpretation of the nine rings is that 
the pattern of ring radii is consistent
with the prediction of the analytic theory 
of CRGs \citep{Appleton}.  Appendix A of 
the present paper gives a brief account of 
the theory and the assumptions on which 
it rests.  The analytic theory predicts that 
the radius $r_i$ of the $i$\textsuperscript{th} ring 
($i = 1, 2, 3, \dots$) 
is proportional to $1/(2 i -1)$.  The ratios of 
successive ring radii are therefore
\begin{equation}
\frac{r_i}{r_{i+1}} = \frac{2 i + 1}{2 i -1} \,.
\label{CRGrat}
\end{equation}
The best fit occurs in case the outermost 
observed ring is $i = 2$.  The $i = 1$ ring, which 
would then have radius 210 kpc, has been searched 
for but has not been found.   The third, fifth and 
sixth columns of Table I list the $i$ labels of 
the rings in the CRG interpretation, the observed 
ring ratios $r_i/r_{i-1}$ and the CRG  predictions 
for those ratios.  The agreement is very good except 
perhaps for the first ratio $r_3/r_2$  which is 
observed to be 0.54 and predicted to be 0.60.     
  
\citet{bull} identifies the smaller galaxy, 
called the impactor, that passed through the 
center of LEDA in the CRG interpretation of its
rings.  Assuming a perpendicular head-on collision, it is 
inferred that the impactor passed through 
LEDA's center 56 Myr ago with a relative speed 
$v_{\rm imp} \simeq 705$~km/s. A trail of gas 
observed between the impactor and the LEDA galaxy 
lends support to the identification of the impactor 
and thus to the CRG interpretation of LEDA's rings.

\section{Critique of the Collisional Ring Interpretation}
In this section we point to difficulties with 
the CRG interpretation of LEDA's nine rings. For completeness and definiteness, Appendix~\ref{app:CRG} describes the analytic theory \citep{Struck,Appleton} on which the CRG interpretation of LEDA's rings \citep{bull} is based.

The analytic theory predicts the ring radii to be 
\begin{equation}
r_i(t) = \frac{\sqrt{2} v_{\rm rot} t}{(2 i -1) \pi} \,,
\label{rira}
\end{equation}
where $t$ is time since the impactor's passage  
through the center of LEDA, and $v_{\rm rot}$
is LEDA's rotation speed.  Since $v_{\rm rot}
\simeq 370$ km/s, the $i=2$ ring expands with 
a speed of 55 km/s. But, for it to have radius 
70 kpc at $t$ = 56 Myr, it would need to expand 
at 1,225 km/s.   This discrepancy by more than a 
factor 20 cannot be remedied by assuming that  
the impact occurred much earlier than 56 Myr ago 
because the density waves and associated rings
fade over a timescale of 100 to 200 Myr
\citep{fade1,fade2,fade3}. Since we need $t\lesssim200$~Myr for collisional rings to exist at all, the speed of the $i=2$ ring must exceed 340~km/s, which is still six times larger than the prediction by the analytic theory.

Fig.~\ref{fig:rvq} shows the radial displacements, according 
to the analytic theory, of test particles 
initially on circular orbits in the disk.  Fig.~\ref{fig:den} 
shows the resulting enhancement of the density 
of test particles as a function of radius.
In both figures, all distances are in units of 
$\frac{\sqrt{2}}{\pi} v_{\rm rot} t$, 
which is the radius of the $i=1$ ring.  The 
amplitude of the radial oscillations is 
proportional to the parameter $\Delta v_r$ 
defined in Eq.~(\ref{delvr}) of Appendix~\ref{app:CRG}.  
We set $\Delta v_r = 60$ km/s so that caustics 
form near the locations of the $i = 2, 3, \dots, 10$ 
rings but not near the location of the  $i=1$
ring. This choice of $\Delta v_r$ may explain 
the non-observation of the $i = 1$ ring.  At 
any rate, $\Delta v_r >$ 55 km/s is needed to 
keep the caustics for $i \geq 2$.  For each 
$i \geq$ 2  there is then a pair of caustics 
separated by a difference $\Delta r_i$ in 
radius.  According to the analytic theory, 
the $i$\textsuperscript{th} collisional ring is located between 
the $i$\textsuperscript{th} pair of caustics at the radius $r_i$ 
where  $\frac{\partial^2 r}{\partial q^2} = 0$. 
The caustics on either side of $r_i$ are taken 
to be the edges of the $i$\textsuperscript{th} ring, so that 
the $i$\textsuperscript{th} ring has width $\Delta r_i$.  For 
illustrative purposes, the horizontal dashed 
line in Fig.~\ref{fig:rvq} shows the radius  $r_3$ of the 
third ring.  In Fig.~\ref{fig:den}, the $r_i$ are indicated 
by vertical dashed lines for $i = 1, 2, 3, 
\dots, 10$.  Figs. 1 and 2 show that $\Delta r_i/r_i$ 
increases with $i$. When $\Delta v_r$ is 
adjusted so that the $i=2$ ring is narrow, 
the $i = 7, 8, 9, 10$ rings overlap each other. 
One would expect therefore the $i = 7, 8, 9, 10$
rings to be blurred. In contrast, the four 
smallest rings in LEDA are observed to be 
distinct.

Finally, we consider the assumptions underlying 
the analytic theory. According to the theory 
the effect of the impactor's passage is 
to give at $t=0$ a sudden impulse to the 
particles in the disk.  However, the condition 
for the validity of this sudden approximation, 
$2 \sqrt{2} v_{\rm rot} \ll v_{\rm imp}$, 
is not satisfied since 
$2 \sqrt{2} v_{\rm rot} \simeq$  1,046 km/s 
whereas $v_{\rm imp} \simeq$ 705 km/s. Although the latter value is inferred in \citet{bull} under idealized assumptions, such
as a head-on and perpendicular collision, the actual impact speed is very unlikely
to satisfy the criterion for a sudden impact.
The analytic theory assumes also that the 
particles in the disk are collisionless.   
This approximation is valid for stars but 
not for gas because the mean free path of 
gas molecules in galactic disks is of order 
parsec and hence too short. When the caustics 
associated with the outermost ring pass a 
given location, the gas particles collide 
with each other there and therefore deviate 
from the behavior ascribed to them in the 
analytic theory. As far as the gas is 
concerned, only the outermost ring is 
motivated. Simulations of the effects of 
the passage of an impactor through the 
center of a disk galaxy in a head-on 
collision typically produce a single 
collisional ring \citep{Appleton,fade1,fade2,fade3}.

\section{Interpretation in terms of caustic
rings of dark matter}

In this section, we propose that the rings 
observed in LEDA are the imprints of caustic 
rings of dark matter on the baryonic matter 
in its disk. Our proposal assumes that the dark matter is axions. We assume further that cold dark matter axions thermalize by their gravitational self-interactions and form a Bose-Einstein condensate, as described in \citet{CABEC,Erken,Banik}. That cold dark matter axions thermalize by their gravitational self-interactions and form a Bose-Einstein condensate is not widely accepted by the community. Although we believe there are excellent reasons to believe in axion thermalization and Bose-Einstein condensation, we state these ingredients here as assumptions to alert the reader to the fact that they may be regarded as speculative.

Caustic rings of dark matter form in galactic 
halos provided the dark matter falls in and 
out of the galactic gravitational potential 
well with large scale vorticity \citep{crdm,
sing}.  Caustics are generic features in the 
distribution of dark matter in galactic halos 
because dark matter is cold and collisionless.  
Caustics form wherever the 3-dimensional sheet 
on which the dark matter particles lie in 
6-dimensional phase space folds back upon 
itself in physical space.  

Present $N$-body simulations of structure 
formation, where the typical particle mass 
is $10^5~M_\odot$,  have inadequate resolution 
to reveal the expected caustics.  When the 
particle mass is $10^5~M_\odot$, the uncertainty 
on particle position due to two-body relaxation 
is of order the halo size.  For a derivation of 
this statement see, for example, the appendix 
of \citet{comet}. The phase space structure 
of galactic halos and the expected caustics are 
therefore washed out in the simulations.
Two-body relaxation is entirely negligible 
for the widely accepted cold dark matter 
candidates such as axions, weakly interacting
dark matter particles (WIMPs), and sterile 
neutrinos \citep{Bertone}.   

Caustic rings of dark matter are circular 
tubes in galactic disks. The dark matter 
density is enhanced at and near the caustics.  
The cross-section of a caustic ring is a section 
of the elliptic umbilic catastrophe ($D_{-4}$) 
\citep{sing}.  It has three cusps one of which 
points away from the galactic center.  The 
caustic ring radii are predicted to be \citep{crdm}
\begin{equation}
r_n \simeq \frac{40~{\rm kpc}}{n}
\left(\frac{v_{\rm rot}}{220~{\rm km/s}}\right)
\left(\frac{j_\mathrm{max}}{0.18}\right)
\label{crr}
\end{equation}
where $n = 1, 2, 3, \dots$, $v_{\rm rot}$ is 
the galaxy's rotation speed, and $j_\mathrm{max}$ is a 
dimensionless measure of its angular momentum. It is related to the galactic spin parameter $\lambda$ \citep{Peebles1,Peebles2} by \citep{case}
\begin{equation}  \label{eq:lambda}
  \lambda \simeq 0.28\,j_\mathrm{max} \,.
\end{equation}
The nominal values appearing in Eq.~(\ref{crr}) 
are those for the Milky Way. The caustic rings are located where the particles falling in along the galactic plane are at their distance of closest approach to the galactic center before falling back out. Eq.~(\ref{crr}) 
was derived using the self-similar model of 
galactic halo formation \citep{FG1,FG2} modified 
to allow for the presence of angular momentum 
and non-radial particle orbits \citep{STW1,STW2,Duffy}. Furthermore, galactic rotation curves were assumed to be flat when both the baryonic and dark matter contributions are included. Actual rotation curves are not exactly flat, of course. Also, one must keep in mind that the caustic rings occur in flows that extend outward to distances of order 10 times the caustic ring radii. Deviations from exact self-similarity on such large scales may be expected. For these reasons the caustic ring radii given in Eq.~(\ref{crr}) should not be taken to be an exact prediction, but rather an approximate prediction to be compared with observations.

Observational evidence has been found in support 
of caustic rings of dark matter with radii 
described by Eq.~(\ref{crr}).  A summary of 
the evidence can be found in \citet{Chakrabarty}. In the Milky Way, where evidence is claimed for caustic rings with $n=2,3$ and $5,6,7,\dots,13$, the deviations from the predictions of Eq.~(\ref{crr}) are of order 3\% \citep{crdm}. In M31, evidence is claimed for the $n=1,2$ and 3 rings, $j_\mathrm{max} \simeq 0.12$ and the deviations from the predictions of Eq.~(\ref{crr}) are also of order 3\%. Eq.~(\ref{eq:lambda}) implies that the spin parameters of the Milky Way and M31 are respectively 0.05 and 0.034. They fall within the expected range, $0.01 \lesssim \lambda \lesssim 0.2$, of galactic spin parameters \citep{Efstathiou1979:spin_parameter,Barnes1987:tidal_torquing}.

A necessary condition for the appearance of 
caustic rings is that the dark matter have 
large scale vorticity.  This requirement allows 
a distinction to be made between axions and 
other dark matter candidates.  Large scale vorticity 
is absent if the dark matter is WIMPs, sterile 
neutrinos or wave dark matter described in terms
of a classical field \citep{Natarajan}.  Axions,
on the other hand, do acquire vorticity because 
they thermalize as a result of their gravitational 
self interactions \citep{CABEC,Erken}. When they 
thermalize, they form a Bose-Einstein condensate, 
meaning that most axions go to the lowest energy 
state available to them through the thermalizing 
interactions.  The lowest energy state for given 
angular momentum is a state of rigid rotation, and
this is a state of large scale vorticity \citep{Banik}.  
It was shown in \citet{case} that Bose-Einstein 
condensation of cold dark matter axions explains 
in detail the occurrence and the properties of 
caustic rings of dark matter, including the pattern 
of ring radii given in Eq.~(\ref{crr}).  More 
recently it was shown that Bose-Einstein condensation 
of cold dark matter axions explains the formation of 
supermassive black holes at cosmic dawn \citep{smbh}.

We now argue that the nine rings observed 
in the LEDA galaxy are the imprints upon 
baryonic matter of caustic rings of dark 
matter.  First, caustic rings of dark matter 
are naturally centered on galaxies because 
they are a property of the galaxies 
themselves.  They do not require an 
accident akin to the passing of an impactor 
through the galactic center in a head-on 
collision.  That LEDA's rings are not exactly 
centered is also easy to understand because 
caustic rings form in flows of dark matter 
extending to ten to twenty times the ring 
radius \citep{Duffy}. Since each flow 
experiences a somewhat different galactic 
gravitational potential, one would not 
expect the caustic rings to be exactly 
centered. Offsets have been inferred from the $n=5$ ring in the Milky Way and the $n=2$ ring in M31 \citep{Chakrabarty}.

Second, the pattern of ring radii predicted
by Eq.~(\ref{crr}) is qualitatively consistent 
with the observations in LEDA.  The fourth 
and seventh columns of Table~\ref{tbl1} give the 
$n$ labels of the nine rings and the predicted
ratios of successive ring radii. We performed a $\chi^2$ fit of the observed ring radii to the predictions of Eq.~(\ref{crr}), with $j_\mathrm{max}$ the only fitting parameter. The errors on the observed radii were taken to be those stated in \citet{bull} and listed in the second column of Table~\ref{tbl1}. When Eq.~(\ref{crr}) is fitted to the observed radii without allowing for theoretical errors, the outcome is $j_\mathrm{max}=0.19$ and $\frac{1}{8}\chi^2 = 1.94$. For comparison, when the observed radii are fitted to the CRG prediction of Eq.~(\ref{CRGrat}) with $r_1$ as the only fitting parameter, $\frac{1}{8}\chi^2 = 0.42$. For the reasons stated above, it is appropriate to allow for theoretical errors. For the latter we took $\delta r^\mathrm{th}_n = \alpha r_n$ where $\alpha$ is a parameter. Assuming the total errors to be $\sigma_n = \sqrt{\delta r_n^2 + (\alpha r_n)^2}$, the fit was performed for different values of $\alpha$. We found $\frac{1}{8} \chi^2 = 1.0$ when $\alpha = 0.077$. So Eq.~(\ref{crr}) is accurate at the level of 7.7\% in the case of LEDA. Table~\ref{tbl1} shows that the theoretically predicted ratios of neighboring ring radii (7\textsuperscript{th} column) are systematically smaller that the observed ratios (5\textsuperscript{th} column), indicating that the rings are spread less widely than predicted by Eq.~(\ref{crr}). The 8\textsuperscript{th} column of Table~\ref{tbl1} shows the caustic ring radii in the fit to the observed radii for $\alpha = 0.077$. Although the fit of Eq.~(\ref{crr}) to LEDA's rings is imperfect, it is still highly significant since the probability that the 8 ring ratios agree with 7.7\% is of order $(0.077)^8 \simeq 1.2\times 10^{-9}$ if the ratios were random numbers.

Finally, caustic rings of dark matter move slowly.
The outward speed of the $n$\textsuperscript{th} ring is 
\begin{equation}
\dot{r}_n \simeq \frac{4}{3 t_0} r_n \,,
\label{speed}
\end{equation}
where $t_0$ is the present age of the Universe.
Caustic rings are plausible sites of new star 
formation because they gravitationally attract 
surrounding baryonic matter \citep{Nat2,Chak2}.  
Bursts of new star formation last 200 to 400 Myr 
\citep{McQuinn}. One expects a minimum spread of 
the star forming regions in the radial direction 
by the amount $\Delta r_n =\dot{r}_n \Delta t$  
where $\Delta t$ is the duration of star formation 
bursts.  Prominent new star formation occurs at 
the locations of the $n = 2$ and 3 rings in LEDA, 
which move with speeds of 3.7 and 2.6 km/s respectively. 
The minimum spread of the star forming regions 
caused by the $n$ = 2 and 3 rings is of order 
1~kpc.  This is consistent with the observations 
in LEDA.  It would be inconsistent with the 
observations in LEDA if the rings were to 
move, say, ten times faster. It has been possible to measure the radial speeds of rings in some galaxies \citep{CRGRadialSpeed1,CRGRadialSpeed2}. A measurement of the radial velocities of the rings in LEDA would provide a powerful test of the CRG and caustic ring interpretations, assuming that the actual ring speeds are measured as distinct from the speed of the stars or gas in the rings.

In conclusion, we have argued that the rings observed in LEDA are the imprint of caustic rings of dark matter on baryonic matter. We do not claim that all galactic rings are the imprints of caustic rings of dark matter, of course. Many observed rings are no doubt collisional. On the other hand, we do claim that the axion dark matter hypothesis implies that all isolated disk galaxies have caustic rings. Under what conditions a caustic ring of dark matter produces a visible imprint on the baryonic matter is not clear at present. The effect of a caustic ring of dark matter on the distribution of stars and gas in galactic disks is discussed in \citet{Nat2,Chak2}. Under what circumstances the enhanced density of gas results in a burst of star formation remains to be investigated. A campaign to search systematically for rings in isolated disk galaxies may possibly reveal that many have multiple rings like those in LEDA.

\begin{acknowledgments}

We are grateful to Laura Blecha, Anthony Gonzalez, 
Abhishek Chattaraj and the members of the UF Particle 
Theory Group for useful discussions and critical feedback.  
This work was supported in part by the U.S. Department of 
Energy under grant DE-SC0022148 at the University of 
Florida.  

\end{acknowledgments}

\begin{table}[htbp]
\caption{The Observed Radii of the Nine Rings Compared to Theoretical Predictions.}
\label{tbl1}

\setlength{\tabcolsep}{15pt}
\begin{center}
\begin{tabular}{|c|c|c|c|c|c|c|c|}
\hline
$r_i$ (pixels) & $\delta r_i$ & $i$ & $n$ & $\frac{r_i}{r_{i-1}}$ & $\frac{2 i -3}{2 i - 1}$ & $\frac{n-1}{n}$ & $\hat{r}_n$ (pixels) \\
\hline
  1390 & 251 &  2 & 1 & \dots & \dots & \dots & 1697 \\
  748  & 39  &  3 & 2 & 0.538 & 0.600 & 0.500 & 848  \\
  533  & 33  &  4 & 3 & 0.713 & 0.714 & 0.667 & 566  \\
  427  & 35  &  5 & 4 & 0.801 & 0.777 & 0.750 & 424  \\
  363  & 31  &  6 & 5 & 0.850 & 0.818 & 0.800 & 339  \\
  321  & 26  &  7 & 6 & 0.884 & 0.846 & 0.833 & 283  \\
  278  & 30  &  8 & 7 & 0.866 & 0.867 & 0.857 & 242  \\
  253  & 25  &  9 & 8 & 0.910 & 0.882 & 0.875 & 212  \\
  202  & 24  & 10 & 9 & 0.798 & 0.894 & 0.889 & 188  \\
\hline
\end{tabular}
\end{center}
\tablecomments{\raggedright$i$ is the label assigned to the rings in the CRG interpretation. $n$ is the label assigned to the rings in the CRDM interpretation. The second column gives the errors on the ring radii. The last column shows the ring radii in the CRDM interpretation fitted to the observed radii as described in the text.}

\end{table}

\clearpage

\appendix

\section{The analytic theory of collisional 
ring galaxies} \label{app:CRG}

This appendix gives a brief account of the analytic 
theory of CRG galaxies in \citet{Struck,Appleton}.
A compact galaxy, called the impactor, falls with speed 
$v_{\rm imp}$ through the center of a disk 
galaxy, called the target galaxy, along the direction 
perpendicular to the disk.  Consider a test particle 
moving on a circular orbit in the disk of the target 
galaxy at galactocentric radius $q$. To obtain Eq.~(\ref{rira}) it is assumed that both the target and impactor galaxies have flat rotation curves. We make this assumption henceforth. A more general case with non-flat rotation curves is discussed in \citet{Struck,Appleton}. Over the course 
of the impactor's passage the test particle is given 
an inward velocity of magnitude  
\begin{equation}
\Delta v_r \sim 2 
\frac{v_{\rm rot}^{\prime~2}}{v_{\rm imp}} \,,
\label{delvr}
\end{equation}
where $v_{\rm rot}^{\prime}$ is the impactor's rotation 
speed. To obtain Eq.~(\ref{delvr}), assume that the 
inward gravitational field due to the impactor has 
magnitude $v_{\rm rot}^{\prime~2}/q$ and that it 
acts over a time of order $2 q/v_{\rm imp}$. As a 
result of the impactor's passage, the test particle 
begins radial oscillations with frequency 
\begin{equation}
\omega_r(q) = \sqrt{2} \frac{v_{\rm rot}}{q} \,,
\label{omr}
\end{equation}
where $v_{\rm rot}$ is the rotation speed of the 
target galaxy.

The CRG analytic theory assumes that the 
particles in the disk all start their radial 
oscillation suddenly at the moment, taken to be 
at time $t = 0$, when the impactor passes through 
the center of the target galaxy. The subsequent
radial positions of the test particles are then 
given by
\begin{equation}
r(q,t) = q - q 
\frac{\Delta v_r}{\sqrt{2} v_{\rm rot}}
\sin\left(\frac{\sqrt{2} v_{\rm rot} t}{q}\right) \,.
\label{rqt}
\end{equation}
This sudden approximation is valid for the test 
particle at $q$ provided the time of passage 
$2 q/v_{\rm imp}$ is short compared 
to $1/\omega_r(q)$, i.e. provided
\begin{equation}
2 \sqrt{2} v_{\rm rot} \ll v_{\rm imp} \,.
\label{impcon}
\end{equation}
The condition is independent of $q$.

\begin{figure}
\begin{center}
\includegraphics[height=110mm]{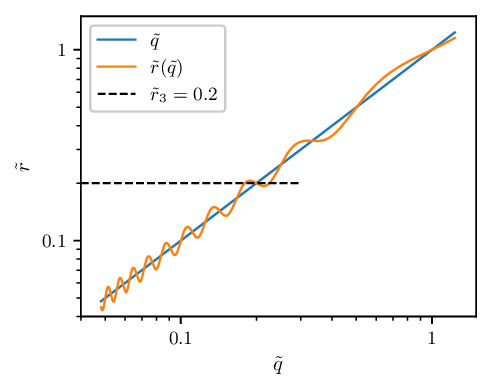}
\caption{Plot of $r(q,t)$ given in Eq.~(\ref{rqt}).
$r$ and $q$ are in units of 
$\frac{\sqrt{2}}{\pi} v_{\rm rot} t$.  The 
horizontal dashed line indicates, for illustrative
purposes, the position of the third ring in the 
analytic theory of collisional ring galaxies.}
\label{fig:rvq}
\end{center}
\end{figure}

\begin{figure}
\begin{center}
\includegraphics[height=110mm]{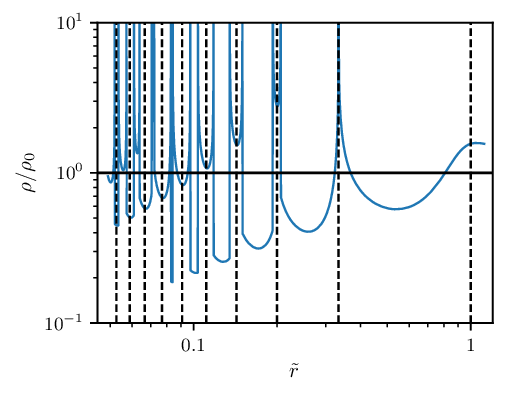}
\vspace{0.3in}
\caption{Plot of the density $\rho(r,t)$ in units of the unperturbed density $\rho_0(q(r,t))$ at the initial location, according to Eqs.~(\ref{denj}) and (\ref{rqt}). $\tilde{r}$ is $r$ in units of $\frac{\sqrt{2}}{\pi} v_{\rm rot} t$. The vertical dashed lines indicate the ring radii $\tilde{r}_i$ for $i = 1, 2, \dots, 10$.}
\label{fig:den}
\end{center}
\end{figure}

Fig.~\ref{fig:rvq} shows the function $r(q,t)$ with $r$ 
and $q$ in units of the length 
$\frac{\sqrt{2}}{\pi} v_{\rm rot} t$.  This 
rescaling removes the dependence both on $t$
and on $v_{\rm rot}$. For sufficiently large
$\Delta v_r$ caustics appear because there 
is more than one value of $q$ for a given $r$.
The density at location $r$ is given by 
\begin{equation}
\rho(r,t) = \frac{1}{r} 
\sum_j \frac{q~\rho_0(q)}{ 
|\frac{\partial r}{\partial q}|} \Bigg|_{q=q_j(r,t)} \,,
\label{denj}
\end{equation}
where the sum is over the solutions $q_j(r,t)$
of $r = r(q,t)$ and $\rho_0(q)$ is the density 
before impact.  The caustics are located where 
$\frac{\partial r}{\partial q} = 0$, i.e. at 
those locations where the density in Eq.~(\ref{denj})
is formally infinite. Fig.~\ref{fig:den} shows the density 
implied by $r(q,t)$ in Fig.~\ref{fig:rvq}. 
The caustics come in pairs. 
The CRG analytic theory does not place 
the rings at the caustics but at the radii where 
\begin{equation}
\frac{\partial^2 r}{\partial q^2} 
= \frac{\sqrt{2} \Delta v_r v_{\rm rot} t^2}{q^3}
\sin\left(\frac{\sqrt{2} v_{\rm rot} t}{q}\right)
= 0
\label{sec}
\end{equation}
with $\frac{\partial r}{\partial q} < 0$, between pairs of caustics. The ring radii are then 
\begin{equation}
r_i = \frac{\sqrt{2} v_{\rm rot} t}{(2 i - 1) \pi} \,,
\label{ri}
\end{equation}
for $i = 1 ,2 ,3, \dots$, and they move at the speeds
\begin{equation}
v_i = \frac{\sqrt{2} v_{\rm rot}}{(2 i - 1) \pi}
\label{vi}
\end{equation}
starting at $r=0$ when $t=0$.

The horizontal dashed line in Fig.~\ref{fig:rvq} shows, for
illustrative purposes, the location of the third 
ring ($i = 3$).  The vertical dashed lines in Fig.~\ref{fig:den} 
show the locations of the first ten rings. Whether 
and how many caustics appear depends on the value 
of $\Delta v_r$ relative to $v_{\rm rot}$. Clearly, 
there are no caustics if $\Delta v_r = 0$. For 
$v_{\rm rot}$ = 370 km/s, the two $i = 1$ caustics 
appear only if $\Delta v_r > 167$ km/s, the two 
$i = 2$ caustics appear only if $\Delta v_r > 55$ 
km/s, and the two $i = 3$ caustics appear only if 
$\Delta v_r > 33$ km/s.  In Figs.~\ref{fig:rvq} and \ref{fig:den}, 
$\Delta v_r$ = 60 km/s was chosen so that there 
are no caustics for $i=1$ but there are caustics
for $i = 2, 3, \dots, 10$, as this best fits the 
CRG interpretation of the nine rings of LEDA.

\newpage

\bibliography{ref}
\bibliographystyle{aasjournalv7}

\newpage

\maxdeadcycles=200

\end{document}